\documentclass[preprint,aps,nofootinbib]{revtex4}
\usepackage[dvips]{graphicx}
\usepackage{amsmath, amsthm, amssymb, mathrsfs}

\newcommand{\alp}{\alpha}
\newcommand{\bta}{\beta}
\newcommand{\gmm}{\gamma}

\newcommand{\lt}{\left}
\newcommand{\rt}{\right}

\newcommand{\ket}[1]{\lvert #1 \rangle}
\newcommand{\bra}[1]{\langle #1 \rvert}
\newcommand{\braket}[2]{\langle #1 \mid #2 \rangle}

\newcommand{\abs}[1]{\lvert #1 \rvert}

\begin{document}
\title{Toward an understanding of entanglement for generalized
n-qubit W-states}
\author{Levon Tamaryan}
\affiliation{Physics Department, Yerevan State University,
Yerevan, 375025, Armenia}
\author{Hungsoo Kim}
\affiliation{The Institute of Basic Science, Kyungnam University,
Masan, 631-701, Korea}
\author{Eylee Jung, Mi-Ra Hwang, DaeKil Park}
\affiliation{Department of Physics, Kyungnam University, Masan,
Korea}
\author{Sayatnova Tamaryan}
\affiliation{Theory Department, Yerevan Physics Institute,
Yerevan, Armenia}

\begin{abstract}
We solve stationarity equations of the geometric measure of
entanglement for multi-qubit W-type states. In this way we compute
analytically the maximal overlap of one-parameter $n$-qubit and
two-parameter four-qubit W-type states and their nearest product
states. Possible extensions to arbitrary W-type states and
geometrical interpretations of these results are discussed in
detail.
\end{abstract}


\maketitle

\section{Introduction}
Entanglement of quantum states~\cite{ami07} plays an important
role in quantum information, computation and communication(QICC).
It is a genuine physical resource for the teleportation
process~\cite{ben,08-mixed} and makes it possible that the quantum
computer outperforms classical one~\cite{vidal03-1,joz}. It also
plays a crucial role in quantum cryptographic
schemes~\cite{ek,fuchs97}. These phenomena have provided the basis
for the development of modern quantum information science.

Quantum entanglement is a rich field of research. A better
understanding of quantum entanglement, of ways it is
characterized, created, detected, stored and manipulated, is
theoretically the most basic task of the current QICC research. In
bipartite case entanglement is relatively well understood, while
in multipartite case even quantifying entanglement of pure states
is a great challenge.

The geometric measure of entanglement can be considered as one of
the most reliable quantifiers of multipartite
entanglement~\cite{shim95,barn,wei03-1}. It depends on $P_{max}$,
the maximal overlap of a given state with the nearest product
state, and is defined by the formula $E_g(\psi) = 1 -
P_{max}$~\cite{wei03-1}. The same overlap $P_{max}$, known also as
the injective tensor norm of $\psi$~\cite{wern}, is the maximal
probability of success in the Grover's search
algorithm~\cite{grover97} when the state $\psi$ is used as an
input state. This relationship between the success probability of
the quantum search algorithm and the amount of entanglement of the
input state allows oneself to define an operational entanglement
measure known as Groverian entanglement~\cite{biham-1,shim04}.

The maximal overlap $P_{max}$ is a useful quantity and has
several practical applications. It has been used to study quantum
phase transitions in spin models~\cite{wei-phase,orusand} and to
quantify the distinguishability of multipartite states by local
means~\cite{local}. Moreover, $P_{max}$ exhibits interesting
connections with entanglement witnesses and can be efficiently
estimated in experiments~\cite{guh-06}. Recently, it has been shown that
the maximal overlap is the largest coefficient of the generalized
Schmidt decomposition and the
nearest product state uniquely defines the factorizable basis of
the decomposition~\cite{hig,gsd-08}.

In spite of its usefulness one obstacle to use $P_{max}$ fully in
quantum information theories is the fact that it is difficult to
compute it analytically for generic states. The usual maximization
method generates a system of nonlinear equations~\cite{wei03-1}.
Thus, it is important to develop a technique for the computation of
$P_{max}$~\cite{jung07-1,wei-sev,sud-geom,maxim,wei-guh}.

Theorem I of Ref.\cite{jung07-1} enables us to compute $P_{max}$
for $n$-qubit pure states by making use of $(n-1)$-qubit reduced
states. In the case of three-qubit states the theorem effectively
changes the nonlinear eigenvalue equations into the linear form.
Owing to this essential simplification $P_{max}$ for the
generalized three-qubit W-state~\cite{w,preeti} was computed
analytically in Ref.\cite{tama07-1}. Furthermore, in
Ref.\cite{tama08-1} $P_{max}$ was found for three-qubit
quadrilateral states with an elegant geometric interpretation.
More recently, based on the analytical results of
Ref.\cite{tama07-1,tama08-1} and the classification of
Ref.\cite{acin00}, $P_{max}$ for various types of three-qubit
states was computed analytically  and expressed in terms of local
unitary(LU) invariants~\cite{jung08-1}.

In general, the calculation of the multi-partite entanglement is
confronted with great difficulties. Furthermore, even if we know
explicit expressions of entanglement measure, the separation of
the applicable domains is also a nontrivial task~\cite{tama08-1}.
Therefore, there is a good reason to consider first some solvable
cases that allow analytic solutions and clear separations of the
validity domains. Later, these results could be extended, either
analytically or numerically, for a wider class of multi-qubit
states. In the light of these ideas we consider one- and
two-parametric $n$-qubit W-type states with $n \geq 4$ in this
paper.

The paper is organized as follows. In Sec. II we clarify our tasks
and notations. In Sec. III we review the calculational tool
introduced in Ref.\cite{jung07-1,tama07-1,tama08-1}  and explain
how the Lagrange multiplier method gives simple solution to the
one-parameter cases. This method is used Sec. IV for the
derivation of $P_{max}$ for one-parameter W-states in $4$-qubit,
$5$-qubit and $6$-qubit systems. In this section the analytical results
are compared with numerical data. In Sec. V based on the analytical
results of the previous section we compute $P_{max}$
for an one-parameter W-state in arbitrary $n$-qubit
system. In Sec. VI we derive $P_{max}$ for two-parameter W-states
in $4$-qubit system by adopting the usual maximization technique. In
Sec.VII we analyze two-parameter results by considering several particular
cases. In Sec. VIII we discuss the possibility of extensions
of the results to arbitrary W states and the existence of
a geometrical interpretation.

\section{Summary of Tasks}

Let $|\psi\rangle$ be a pure state of an $n$-party system ${\cal H} =
{\cal H}_1\otimes{\cal H}_2\otimes\cdots\otimes{\cal H}_n$ , where
the dimensions of the individual state spaces ${\cal H}_k$ are
finite but otherwise arbitrary. The maximal overlap of $|\psi\rangle$ is
given by
\begin{equation}
\label{pmax1} P_{max} (\psi) \equiv \max_{|q_1\rangle \cdots
|q_n\rangle} |\langle q_1 |\langle q_2| \cdots \langle q_n| \psi
\rangle|^2,
\end{equation}
where the maximum is taken over all single-system normalized state
vectors $|q_k\rangle\in{\cal H}_k$, and it is understood that
$|\psi\rangle$ is normalized.

Let us consider now $n$-qubit W-type state
\begin{equation}
\label{wn-state} |W_n\rangle = a_1 |10\cdots 0\rangle + a_2
|010\cdots 0\rangle + \cdots + a_n
                  |0\cdots 01\rangle,
\end{equation}
where the labels within each ket refer to qubits $1,2,\cdots,n$  in
that order.

In this paper we will compute analytically $P_{max}$ in the
following two cases:

1)for the one-parametric $|W_n\rangle$ when $a_1 = \cdots =
a_{n-1}\equiv a$ and $a_n \equiv q$

2)for the two-parametric $|W_4 \rangle$  when $a_1= a, \; a_2 = b,
\; a_3 = a_4 = q$.

To ensure the calculational validity we use the result of \cite{shim04},
which has shown that $P_{max} = (1 - 1/n)^{n-1}$ when
$a_1 = a_2 = \cdots = a_n$. Thus, the final results of the one-parametric
case should agree with the following:

$\bullet$ If $a = q = 1 / \sqrt{n}$, then $P_{max}$ should be
equal to $(1 - 1/n)^{n-1}$.

$\bullet$ If $q=0$, then $|W_n\rangle$ becomes $|W_{n-1}\rangle
\otimes |0\rangle$ and, as a result, $P_{max}$ should be equal to
$(1 - 1 / (n-1))^{n-2}$.

For the two-parametric case $P_{max}(W_4)$ should have
a correct limit when either
$a$ or $b$ vanishes. At $a=0$ we have $|W_4\rangle =
|0\rangle\otimes |W_3\rangle$ and thus the maximal overlap should
be expressed in terms of the circumradius of the isosceles
triangle with sides $b,q,q$~\cite{tama07-1}.

\section{Calculation Tool}

For a pure state of two qubits $P_{max}$ is given by
\begin{equation}
\label{2q} P_{max} = \frac{1}{2} \left[1 + \sqrt{1 - 4 \det
\rho^A} \right],
\end{equation}
where $\rho^A$ is reduced density matrix, {i.e.} $\mbox{Tr}_B
\rho^{AB}$. Therefore, the Bell (and their LU-equivalent)
states have the minimal ($P_{max} = 1/2$) while product states have the
maximal ($P_{max}=1$) overlap.

The explicit dependence of $P_{max}$ on state parameters for the
generalized three-qubit W-state
\begin{equation}
\label{w3-state} |W_3\rangle = a_1 |100\rangle + a_2 |010\rangle +
a_3 |001\rangle
\end{equation}
was computed in Ref.\cite{tama07-1}. In order to express
explicitly $P_{max}(W_3)$  in terms of state parameters, we define
a set $\{\alpha, \beta, \gamma\}$ as the set $\{a_1, a_2, a_3\}$
in decreasing order. Then $P_{max}$ for the generalized W-state
can be expressed in a form
\begin{eqnarray}
\label{pmaxw3} P_{max} (W_3) = \left\{       \begin{array}{ll}
                  4 R_W^2       &  \hspace{1.0cm}
                                        \mbox{when $\alpha^2 \leq \beta^2 + \gamma^2$}  \\
                    \alpha^2    & \hspace{1.0cm}
                                         \mbox{when $\alpha^2 \geq \beta^2 + \gamma^2$}
                        \end{array}
                        \right.
\end{eqnarray}
where $R_W$ is the circumradius of the triangle with sides $a_1$,
$a_2$, $a_3$. Similar calculation procedure can be applied to the
$3$-qubit quadrilateral state. It has been shown in Ref.\cite{tama08-1}
that for this case $P_{max}$ is expressed in terms of the circumradius of
a convex quadrangle.
These two separate results strongly suggest
that $P_{max}$ for an arbitrary pure state has its own geometrical
meaning. If we are able to know this meaning completely, then our
understanding on the multipartite entanglement would be greatly
enhanced.

Now, we briefly review how to derive the analytic result
(\ref{pmaxw3}) because it plays crucial role in next two sections.
In Ref.\cite{tama07-1} $P_{max}$ for $3$-qubit state is expressed
as
\begin{equation}
\label{3-pmax-1} P_{max} = \frac{1}{4} \max_{|\vec{s}_1| =
|\vec{s}_2| = 1} \left[1 + \vec{s}_1 \cdot \vec{r}_1 + \vec{s}_2
\cdot \vec{r}_2 + g_{ij} s_{1i} s_{2 j}
                                                                   \right]
\end{equation}
where $\vec{s}_1$ and $\vec{s}_2$ are Bloch vectors of the
single-qubit states. In Eq.(\ref{3-pmax-1}) $\vec{r}_1 = Tr[\rho^A
\vec{\sigma}]$, $\vec{r}_2 = Tr[\rho^B \vec{\sigma}]$ and $g_{ij}
= Tr[\rho^{AB} \sigma_i \otimes \sigma_j]$, where $\rho^A$,
$\rho^B$ and $\rho^{AB}$ are appropriate partial traces of
$\rho^{ABC} \equiv |W_3\rangle \langle W_3|$ and $\sigma_i$ are
usual Pauli matrices. The explicit expressions of $\vec{r}_1$,
$\vec{r}_2$ and $g_{ij}$ are given in Ref.\cite{tama07-1}. Due to
maximization over $\vec{s}_1$ and $\vec{s}_2$ in
Eq.(\ref{3-pmax-1}) we can compute $\vec{s}_1$ and $\vec{s}_2$ by
solving the Lagrange multiplier equations
\begin{equation}
\label{multiplier1} \vec{r}_1 + g \vec{s}_2 = \lambda_1 \vec{s}_1,
\hspace{2.0cm} \vec{r}_2 + g^T \vec{s}_1 = \lambda_2 \vec{s}_2,
\end{equation}
where $\lambda_1$ and $\lambda_2$ are Lagrange multiplier
constants. Now, we let $s_{1y} = s_{2y} = 0$ for simplicity,
because those give only irrelevant overall phase factor to
$\langle q_1 |\langle q_2| \langle q_3| W_3\rangle$. After
eliminating the Lagrange multiplier constants, one can show that
Eq.(\ref{multiplier1}) reduces to two equations. Examining these
two remaining equations, one can show that $\vec{s}_1$ and
$\vec{s}_2$ have a following relation to each other:
\begin{equation}
\label{rela1} \vec{s}_1 (a_1, a_2, a_3) = \vec{s}_2 (a_2, a_1,
a_3).
\end{equation}
Using this relation, one can combine these two equations into
single one expressed in terms of solely $s_{1z}$ in a final form
\begin{equation}
\label{single-3} \frac{\sqrt{1 - s_{1z}^2 (a_1, a_2, a_3)}}{s_{1z}
(a_1, a_2, a_3)} = \frac{\omega \sqrt{1 - s_{1z}^2 (a_2, a_1,
a_3)}}{r_1 - r_3 s_{1z} (a_2, a_1, a_3)}
\end{equation}
where $r_1 = a_2^2 + a_3^2 - a_1^2$, $r_2 = a_1^2 + a_3^2 -
a_2^2$, $r_3 = a_1^2 + a_2^2 - a_3^2$ and $\omega = 2 a_1 a_2$.
Defining $a_1 = a_2 \equiv a$ and $a_3 \equiv q$ again, one can
solve Eq.(\ref{single-3}) easily in a form
\begin{eqnarray}
\label{3-solu} & &s_{1z} = s_{2z} = \frac{r_1}{\omega + r_3} =
\frac{q^2}{4 a^2 - q^2}
                                                                \\  \nonumber
& &s_{1x} = s_{2x} = \sqrt{1 - s_{1z}^2} = \frac{2 \sqrt{2} a}{4
a^2 - q^2}\sqrt{2 a^2 - q^2}.
\end{eqnarray}
Inserting Eq.(\ref{3-solu}) into Eq.(\ref{3-pmax-1}), one can
compute $P_{max}$ for $|W_3\rangle$ with $a_1 = a_2 = a$ and $s_3
= q$, whose final expression is simply
\begin{equation}
\label{3-pmax-aaq} P_{max} = \frac{(1 - q^2)^2}{2 - 3 q^2}.
\end{equation}
Eq.(\ref{3-pmax-aaq}) is consistent with Eq.(\ref{pmaxw3}) when
$q^2 \leq 2 a^2$. When $q=0$, Eq.(\ref{3-pmax-aaq}) gives $P_{max}
= 1/2$ which corresponds to that of $2$-qubit EPR state. When
$q=1/\sqrt{3}$, Eq.(\ref{3-pmax-aaq}) gives $P_{max} = 4/9$, which
is also consistent with the result of Ref.\cite{shim04}.

\section{Four, five and six qubit W-type states: one-parametric cases}

The method described in the previous section may enable us to
compute $P_{max}$ of four-qubit W-type states. For the case of
arbitrary four-qubit systems $P_{max}$ can be represented in a
form
\begin{eqnarray}
\label{4-pmax-1}
 & &P_{max} = \frac{1}{8} \max_{|\vec{s}_1| =
|\vec{s}_2| = |\vec{s}_3| = 1} \bigg[ 1 + \vec{s}_1 \cdot
\vec{r}_1 + \vec{s}_2 \cdot \vec{r}_2 + \vec{s}_3 \cdot \vec{r}_3
\\ \nonumber
 & &+ s_{1i} s_{2j} g_{ij}^{(3)}+ s_{1i} s_{3j}
g_{ij}^{(2)} + s_{2i} s_{3j} g_{ij}^{(1)} + s_{i1} s_{2j} s_{3k}
h_{ijk}
 \bigg],
\end{eqnarray}
where
\begin{eqnarray}
\label{4-def-gen}
 & &\vec{r}_1 = Tr[\rho^A \vec{\sigma}],
\hspace{.5cm} \vec{r}_2 = Tr[\rho^B \vec{\sigma}],\hspace{.5cm}
\vec{r}_3 = Tr[\rho^C \vec{\sigma}],\\  \nonumber
 & &g_{ij}^{(3)}
= Tr[\rho^{AB} \sigma_i \otimes \sigma_j], \; g_{ij}^{(2)} =
Tr[\rho^{AC} \sigma_i \otimes \sigma],\; g_{ij}^{(1)} =
Tr[\rho^{BC} \sigma_i \otimes \sigma_j]\\\nonumber
 & &h_{ijk} = Tr [\rho^{ABC} \sigma_i \otimes \sigma_j \otimes
 \sigma_k].
\end{eqnarray}

For the case of the generalized four-qubit W-state all
vectors $\vec{r_k}$ are collinear, all matrices $g^{(k)}$ are
diagonal and the vectors $\vec{r_k}$ are eigenvectors of the
matrices $g^{(k)}$ as following:
\begin{equation}
\label{4-def-1}
 \vec{r}_k =(0, 0, r_k),\quad g_{ij}^{(k)} = \left(
\begin{array}{ccc}
            \omega_k  &  0  &  0   \\
            0  &  \omega_k  &  0   \\
            0  &  0  &  -\tilde{r}_k
                          \end{array}             \right),
\quad k=1,2,3.
\end{equation}
In Eq.(\ref{4-def-1}) we defined various quantities as following:
\begin{eqnarray}
\label{4-def-2}
 & &r_k =a_1^2+ a_2^2 + a_3^2 + a_4^2 - 2a_k^2,
\quad\omega_1 = 2 a_2 a_3,\; \omega_2 = 2 a_1 a_3, \; \omega_3 = 2
a_1 a_2.
                                   \\\nonumber
 & & \tilde{r}_1=a_2^2 + a_3^2 - a_1^2 - a_4^2,
\; \tilde{r}_2 = a_1^2 + a_3^2 - a_2^2 - a_4^2,\;
 \tilde{r}_3 = a_1^2 + a_2^2 - a_3^2 - a_4^2.  \quad
\end{eqnarray}
In addition, the non-vanishing components of $h_{ijk}$ are
\begin{equation}
\label{4-def-4} h_{113} = h_{223} = \omega_3 \hspace{1.0cm}
h_{131} = h_{232} = \omega_2 \hspace{1.0cm} h_{311} = h_{322} =
\omega_1 \hspace{1.0cm} h_{333} = - r_4.
\end{equation}
Due to the maximization in Eq.(\ref{4-pmax-1}) the Bloch vectors
should satisfy the following Lagrange multiplier equations:
\begin{eqnarray}
\label{4q-multiplier-1} & &r_{1i} + g_{ij}^{(3)} s_{2j} +
g_{ij}^{(2)} s_{3j} + h_{ijk} s_{2j} s_{3k}
                                                           = \Lambda_1 s_{1i}
                                                              \\   \nonumber
& &r_{2i} + g_{ji}^{(3)} s_{1j} + g_{ij}^{(1)} s_{3j} + h_{kij}
s_{1k} s_{3j}
                                                           = \Lambda_2 s_{2i}
                                                              \\   \nonumber
& &r_{3i} + g_{ji}^{(2)} s_{1j} + g_{ji}^{(1)} s_{2j} + h_{jki}
s_{1j} s_{2k}
                                                           = \Lambda_3 s_{3i}.
\end{eqnarray}
Now we put $s_{1y} = s_{2y} = s_{3y} = 0$ as before. After
removing the Lagrange multiplier constants $\Lambda_1$,
$\Lambda_2$ and $\Lambda_3$, one can show that
Eq.(\ref{4q-multiplier-1}) reduce to the following three
equations:
\begin{eqnarray}
\label{4q-multiplier-2} & & s_{1x} \left[ r_1 - \tilde{r}_3 s_{2z}
- \tilde{r}_2 s_{3z} + \omega_1 s_{2x} s_{3x} - r_4 s_{2z} s_{3z}
\right]
 = s_{1z} \left[\omega_2 s_{3x} (1 + s_{2z})
+ \omega_3 s_{2x} (1 + s_{3z}) \right]
                                                                  \\   \nonumber
& & s_{2x} \left[ r_2 - \tilde{r}_3 s_{1z} - \tilde{r}_1 s_{3z} +
\omega_2 s_{1x} s_{3x} - r_4 s_{1z} s_{3z} \right] = s_{2z}
\left[\omega_1 s_{3x} (1 + s_{1z}) + \omega_3 s_{1x} (1 + s_{3z})
\right]
                                                                  \\   \nonumber
& &s_{3x} \left[ r_3 - \tilde{r}_1 s_{2z} - \tilde{r}_2 s_{1z} +
\omega_3 s_{1x} s_{2x} - r_4 s_{1z} s_{2z} \right] = s_{3z}
\left[\omega_2 s_{1x} (1 + s_{2z}) + \omega_1 s_{2x} (1 + s_{1z})
\right].
\end{eqnarray}
Eq.(\ref{4q-multiplier-2}) implies that the Bloch vectors have
the following symmetries:
\begin{eqnarray}
\label{4q-symmetry1} & & \vec{s}_1 (a_1, a_2, a_3, a_4) =
\vec{s}_2 (a_2, a_1, a_3, a_4) = \vec{s}_3 (a_3, a_2, a_1, a_4)\\
\nonumber
 & & \vec{s}_1 (a_1, a_2, a_3, a_4) = \vec{s}_1 (a_1,
a_3, a_2, a_4) \\\nonumber
 & & \vec{s}_2 (a_1, a_2, a_3, a_4) = \vec{s}_2
(a_3, a_2, a_1, a_4)   \\  \nonumber
 & & \vec{s}_3 (a_1, a_2, a_3,
a_4) = \vec{s}_3 (a_2, a_1, a_3, a_4).
\end{eqnarray}
Therefore, one can compute all Bloch vectors if one of them is
known. Using the symmetries (\ref{4q-symmetry1}), we can make
single equation from Eq.(\ref{4q-multiplier-2}) which is expressed
in terms of $s_{1z}$ only in a form
\begin{equation}
\label{4q-single} \frac{s_{1x} (a_1, a_2, a_3, a_4)} {s_{1z} (a_1,
a_2, a_3, a_4)} = \frac{P (a_1, a_2, a_3, a_4)}{Q (a_1, a_2, a_3,
a_4)}
\end{equation}
where
\begin{eqnarray*}
\label{PandQ} & &P (a_1, a_2, a_3, a_4) = \omega_2 \sqrt{1 -
s_{1z}^2 (a_3, a_2, a_1, a_4)} \left[1 + s_{1z} (a_2, a_1, a_3,
a_4)
                                                                       \right]
                                                                       \\  \nonumber
& &\hspace{2.8cm}+ \omega_3 \sqrt{1 - s_{1z}^2 (a_2, a_1, a_3,
a_4)} \left[1 + s_{1z} (a_3, a_2, a_1, a_4)
                                                                        \right]
                                                                    \\   \nonumber
& &Q (a_1, a_2, a_3, a_4)
 =
r_1 - \tilde{r}_3 s_{1z} (a_2, a_1, a_3, a_4) - \tilde{r}_2 s_{1z}
(a_3, a_2, a_1, a_4)
                                                                    \\   \nonumber
& & \hspace{2.8cm}+ \omega_1 \sqrt{1 - s_{1z}^2 (a_2, a_1, a_3,
a_4)}
           \sqrt{1 - s_{1z}^2 (a_3, a_2, a_1, a_4)}\\\nonumber
 & & \hspace{2.8cm}- r_4 s_{1z} (a_2, a_1, a_3, a_4) s_{1z} (a_3, a_2, a_1, a_4).
\end{eqnarray*}
Defining $a_1 = a_2 = a_3 \equiv a$ and $a_4 \equiv q$, one can
solve Eq.(\ref{4q-single}) easily. The final expressions of
solutions are
\begin{eqnarray}
\label{4q-solu-1} & &s_{1z} = s_{2z} = s_{3z} = \frac{1}{9 a^2 -
q^2}
                                                          \\   \nonumber
& &s_{1x} = s_{2x} = s_{3x} = \sqrt{1 - s_{1z}^2} = \frac{2
\sqrt{6} a}{9 a^2 - q^2} \sqrt{3 a^2 - q^2}.
\end{eqnarray}
Inserting Eq.(\ref{4q-solu-1}) into Eq.(\ref{4-pmax-1}), one can
compute $P_{max}$ for $|W_4\rangle$ with $a_1 = a_2 = a_3 \equiv
a$ and $a_4 \equiv q$ whose final expression is
\begin{equation}
\label{4-pmax-aaaq} P_{max} = \frac{2^2 (1 - q^2)^3}{(3 - 4
q^2)^2}.
\end{equation}
Eq.(\ref{4q-solu-1}) implies that $P_{max}$ in
Eq.(\ref{4-pmax-aaaq}) is valid when $q^2 \leq 3 a^2$. When $q =
0$, $P_{max}$ becomes $4/9$ as expected. When $q = 1/2$, $P_{max}$
becomes $27 / 64$, which is in agreement with the result of
Ref.\cite{shim04}.

One can repeat the calculation for $|W_5\rangle$ with $a_1 = a_2 =
a_3 = a_4 \equiv a$ and $a_5 = q$. Then the final expression of
$P_{max}$ becomes
\begin{equation}
\label{5-pmax-aq} P_{max} = \frac{3^3 (1 - q^2)^4}{(4 - 5 q^2)^3}.
\end{equation}
When $q=0$, $P_{max}$ reduces to $27/64$ as expected. When $q = 1
/ \sqrt{5}$, $P_{max}$ reduces to $(4/5)^4$. By the same way
$P_{max}$ for $|W_6\rangle$ can be written as
\begin{equation}
\label{6-pmax-aq} P_{max} = \frac{4^4 (1 - q^2)^5}{(5 - 6 q^2)^4}.
\end{equation}

\begin{figure}[ht!]
\begin{center}
\includegraphics[height=6cm]{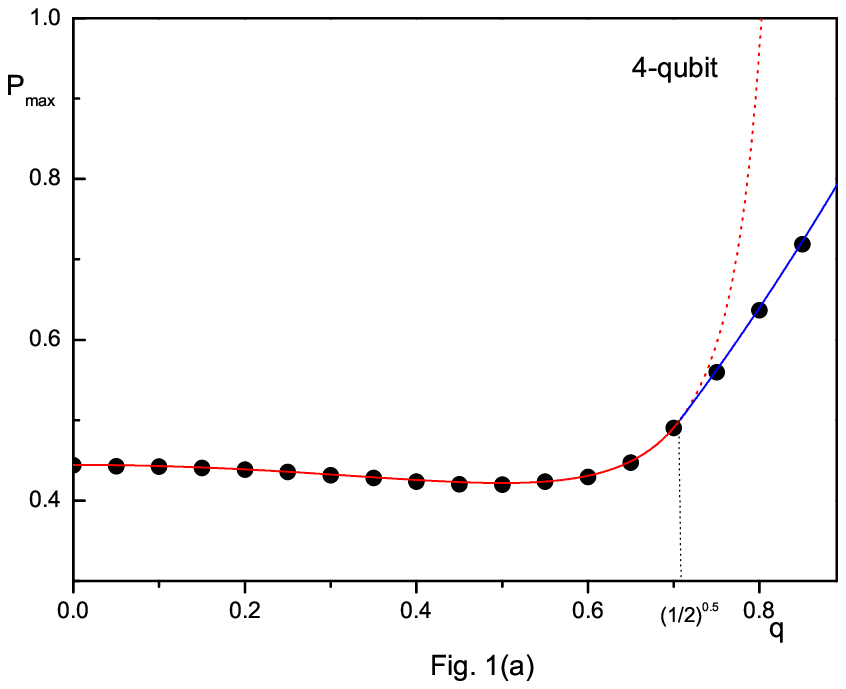}
\includegraphics[height=6cm]{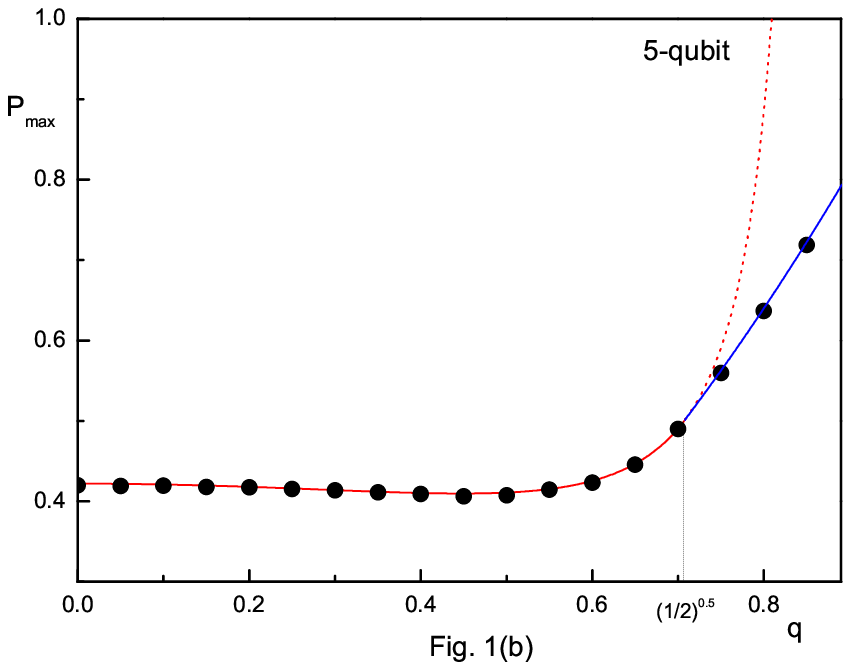}
\includegraphics[height=6cm]{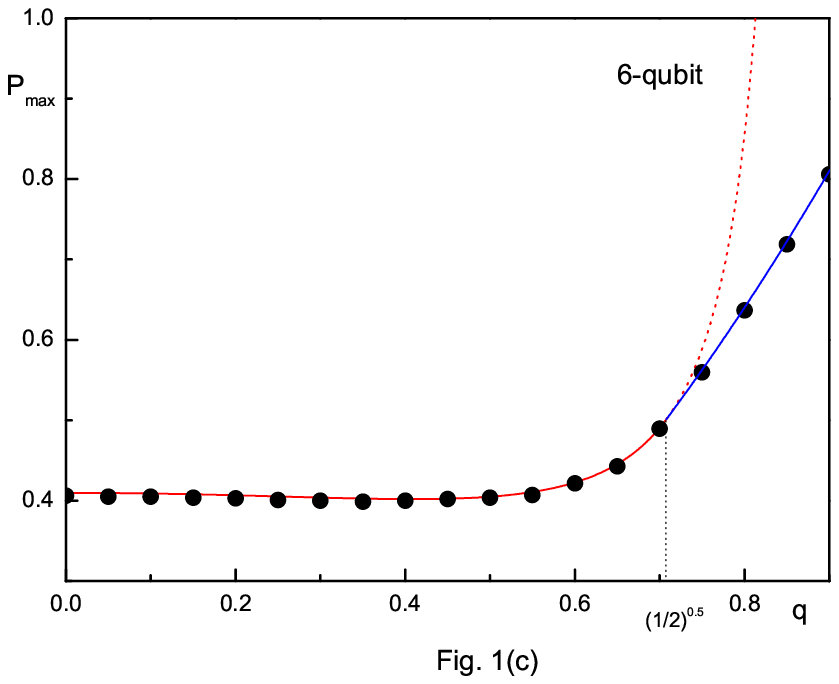}
\caption[fig1]{Plot of $q$-dependence of $P_{max}$ for $4$-qubit
(Fig. 1(a)), $5$-qubit (Fig. 1(b)), and $6$-qubit (Fig. 1(c)). The
black dots are numerical data of $P_{max}$. The red solid lines
are result of Eq.(\ref{n-pmax-aq}) in the applicable domain, $0
\leq q \leq 1 / \sqrt{2}$. The red dotted lines are result of
Eq.(\ref{n-pmax-aq}) outside the applicable domain. The blue solid
lines are plot of $\mbox{max} (a^2, q^2) = q^2$ outside the
applicable domain. This figures strongly suggest that $P_{max}$
for $|W_n\rangle$ is Eq.(\ref{n-pmax-aq}) when $q \leq 1/\sqrt{2}$
and $\mbox{max} (a^2, q^2) = q^2$ when $q \geq 1/\sqrt{2}$. }
\end{center}
\end{figure}

Fig. 1 is a plot of $q$-dependence of $P_{max}$ for $|W_4\rangle$,
$|W_5\rangle$ and $|W_6\rangle$. The black dots are numerical data
computed by the numerical technique exploited in
Ref.\cite{shim04}. The red solid and red dotted lines are
Eq.(\ref{4-pmax-aaaq}), Eq.(\ref{5-pmax-aq}) and
Eq.(\ref{6-pmax-aq}) when $q \leq 1 / \sqrt{2}$ and $q \geq 1 /
\sqrt{2}$ respectively. As expected the numerical data are in
perfect agreement with Eq.(\ref{4-pmax-aaaq}),
Eq.(\ref{5-pmax-aq}) and Eq.(\ref{6-pmax-aq}) in the applicable
domain, {\it i.e.} $q^2 \leq (n-1)  a^2$ for $|W_n\rangle$.
Outside the applicable domain ($q^2 \geq 1 / \sqrt{2}$) the
numerical data are in disagreement with these equations.

\section{General multi-qubit W-type states: one-parametric cases}

From Eq.(\ref{3-pmax-aaq}), (\ref{4-pmax-aaaq}), (\ref{5-pmax-aq})
and (\ref{6-pmax-aq}) one can guess that $P_{max}$ for $W_n$ is
$(a_1=\cdots=a_{n-1}\equiv a,\,a_n\equiv q)$
\begin{equation}
\label{n-pmax-aq} P_{max}(n,q) = (1 - q^2)^{n-1} \left(
\frac{n-2}{(n-1) - n q^2} \right)^{n-2}.
\end{equation}
Using this result, one can straightforwardly construct the nearest
product state to $|W_n\rangle$. After some algebra, when $q^2 \leq
(n-1) a^2$, one can show that the analytic expression of the
nearest product state is $|q_1\rangle \otimes|q_2\rangle\otimes
\cdots \otimes |q_n\rangle$, where
\begin{eqnarray}
\label{nearest} & &|q_1\rangle = \cdots = |q_{n-1}\rangle =
                                                \\  \nonumber
& & \hspace{1.0cm} \frac{1}{\sqrt{(n-1)^2 a^2 - q^2}}
\left[\sqrt{(n-1) (n-2)} a |0\rangle + \sqrt{(n-1) a^2 - q^2}
e^{i\varphi} |1\rangle \right]
                                                                        \\  \nonumber
& &|q_n\rangle = \frac{1}{\sqrt{(n-1)^2 a^2 - q^2}}
\left[\sqrt{(n-1)^2 a^2 - (n-1)q^2} |0\rangle + \sqrt{n-2} q
e^{i\varphi} |1\rangle \right]
\end{eqnarray}
and $\varphi$ is an arbitrary phase factor. When $q^2 \geq (n-1)
a^2$, the nearest product state, of course, becomes $|0 \cdots 0
1\rangle$.

Now, we present a simple proof for both equations (\ref{n-pmax-aq})
and (\ref{nearest}). It is easy to check 
\begin{equation}\label{proof}
\langle q_1q_2\cdots
q_{n-1}|W_n\rangle=e^{-i\varphi}\sqrt{P_{max}}|q_n\rangle,
\quad\langle q_2q_3\cdots q_{n-1}
q_n|W_n\rangle=e^{-i\varphi}\sqrt{P_{max}}|q_1\rangle.
\end{equation}
The second equation in (\ref{proof}) is invariant under the
permutations $(q_1\leftrightarrow q_j, j=2,3,\cdots n-1)$. Thus,
the product state satisfies the stationarity equations of
Ref.\cite{wei03-1} and consequently, is the nearest separable
state. Accordingly, $\sqrt{P_{max}}$ is the injective tensor norm
of $|W_n\rangle$.

When $q=0$ and $q = 1/\sqrt{n}$, $P_{max}$ reduces to $(1 - 1 /
(n-1))^{n-2}$ and $(1 - 1 / n)^{n-1}$ respectively. Thus,
Eq.(\ref{n-pmax-aq}) is perfectly in agreement with the result of
Ref.\cite{shim04}. Another interesting point in
Eq.(\ref{n-pmax-aq}) is that $P_{max}$ becomes $1/2$ regardless of
$n$ when $q = 1/\sqrt{2}$, the boundary of the applicable domain.
This makes us conjecture that outside the applicable domain
$P_{max}$ becomes $\mbox{max} (a^2, q^2) = q^2$ like $3$-qubit
case. The blue solid lines in Fig. 1 are plot of $q^2$ at the domain $q \geq
1 / \sqrt{2}$. As we conjecture, the blue lines are perfectly in
agreement of numerical data.

Another consequence of Eq.(\ref{n-pmax-aq}) is the entanglement
witness $\hat{W_n}$ for an one-parametric W-type state. Its
construction is straightforward as following form:

\begin{equation}\label{witness-1}
\hat{W_n}=P_{max}(n,q) \openone -|W_n(q)\rangle\langle W_n(q)|,
\end{equation}
where $\openone$ is a unit matrix. Obviously one can show

\begin{equation}\label{witness-2}
Tr  \left(\hat{W_n}|W_n(q)\rangle\langle W_n(q)|\right)<0,\quad Tr
\left(\hat{W_n}\rho_0\right) \geq 0,
\end{equation}
where $\rho_0$ is any separable state. Thus, $\hat{W_n}$ is an
entanglement witness and allows an experimental detection of the
multipartite entanglement.


\section{Four-qubit W state: two-parametric cases}

In this section we will compute $P_{max}$ for the two-parametric
$|W_4 \rangle$ given by
\begin{equation}
   \ket{W_4}
      =
         a\ket{1000} + b\ket{0100} + q\ket{0010} + q\ket{0001}.
\end{equation}
It seems to be difficult to apply the Lagrange multiplier method
directly due to their non-trivial nonlinearity. Thus, we will adopt
the usual maximization method.

The maximum overlap probability $P_{max}$ is
\begin{equation}
   \displaystyle
   P_{max}
      =
         \max_{\ket{q_1}\ket{q_2}\ket{q}}
         \abs{\bra{q_1}\bra{q_2}\bra{q}\braket{q}{W_4}}^2.
\end{equation}
Now we define the 1-qubit states as $\ket{q_1} = \alpha_0\ket{0} +
\alpha_1\ket{1}$, $\ket{q_2} = \beta_0\ket{0} + \beta_1\ket{1}$
and $\ket{q} = \gamma_0\ket{0} + \gamma_1\ket{1}$. For simplicity,
we are assuming that all coefficients are real and positive. Then,
$P_{max}$ becomes
\begin{equation}
\label{hs1}
   \begin{array}{l}
      \displaystyle
      P_{max}
         =
            \max_{\alp_0,\bta_0,\gmm_0}
            \gmm_0^2
            \lt(
                  a\bta_0\gmm_0\sqrt{1 - \alp_0^2}
                + b\alp_0\gmm_0\sqrt{1 - \bta_0^2}
                + 2q\alp_0\bta_0\sqrt{1 - \gmm_0^2}
            \rt)^2.
   \end{array}
\end{equation}
Since the maximum value is determined at extremum point, it is useful
if the extremum conditions are derived. This is achieved by differentiating
Eq.(\ref{hs1}), which leads to
\begin{equation}
   \begin{array}{l}
      \displaystyle
         b\gmm_0\sqrt{1 - \bta_0^2} + 2q\bta_0\sqrt{1 - \gmm_0^2}
            =
               a\bta_0\gmm_0\frac{\alp_0}{\sqrt{1 - \alp_0^2}}
                                                                                    \\
      \displaystyle
         a\gmm_0\sqrt{1 - \alp_0^2} + 2q\alp_0\sqrt{1 - \gmm_0^2}
            =
               b\alp_0\gmm_0\frac{\bta_0}{\sqrt{1 - \bta_0^2}}
                                                                                    \\
      \displaystyle
         a\bta_0\gmm_0\sqrt{1 - \alp_0^2} + b\alp_0\gmm_0^2\sqrt{1 - \bta_0^2}
       + q\alp_0\bta_0\sqrt{1 - \gmm_0^2}
            =
               q\alp_0\bta_0\frac{\gmm_0^2}{1 - \gmm_0^2}.
   \end{array}
\end{equation}

One can solve the equations by separating $\alp_0$ from
$\bta_0,\;\gmm_0$, i.e.,
\begin{equation}
   \begin{array}{l}
      \displaystyle
         \frac{\alp_0}{\sqrt{1 - \alp_0^2}}
            =
               \frac{b}{a}\frac{\sqrt{1 - \bta_0^2}}{\bta_0}
             + \frac{2q}{a}\frac{1 - \gmm_0^2}{\gmm_0}
                                                                                    \\
      \scriptscriptstyle
                                                                                    \\
      \displaystyle
         \frac{\sqrt{1 - \alp_0^2}}{\alp_0}
            =
               \frac{b}{a}\frac{\bta_0}{\sqrt{1 - \bta_0^2}}
             - \frac{2q}{a}\frac{\sqrt{1 - \gmm_0^2}}{\gmm_0}
                                                                                    \\
      \scriptscriptstyle
                                                                                    \\
      \displaystyle
         \frac{\sqrt{1 - \alp_0^2}}{\alp_0}
            =
               \frac{q}{a}\frac{\gmm_0}{\sqrt{1 - \gmm_0}}
             - \frac{q}{a}\frac{\sqrt{1 - \gmm_0^2}}{\gmm_0}
             - \frac{b}{a}\frac{\sqrt{1 - \bta_0^2}}{\bta_0}
   \end{array}
\end{equation}
and one can get the solutions for $\bta_0$ and $\gmm_0$ as
follows:
\begin{eqnarray}
& &
      \bta_0^2
         =
            \frac{3}{2}
          - \frac{4q^2 - a^2 + b^2}{4q^2}\gmm_0^2
                                                                              \\  \nonumber
& &
      \gmm_0^2
         =
            \frac
            {
                  4q^2(4q^2 - a^2 - b^2)
              - 2q^2\sqrt{(4q^2 - a^2 - b^2)^2 + 12a^2b^2}
            }
            {(4q^2 + b^2 - a^2)^2 - 16q^2b^2}.
\end{eqnarray}
The solution for $\alp_0$ is obtained by separating $\bta_0$:
\begin{eqnarray}
      \alp_0^2
         =
            \frac{3}{2} - \frac{4q^2 + a^2 - b^2}{4 q^2}\gmm_0^2.
\end{eqnarray}
Inserting these extremum solution in $P_{max}$ and rationalizing
denominator, one gets
\begin{equation}
\label{abqq}
   \begin{array}{l}
      P_{max} = \frac
                     {
                        2q^4
                        \lt[
                              (4q^2 - a^2 - b^2)
                              \lt\{ (4q^2 - a^2 - b^2)^2 - 36a^2b^2 \rt\}
                          + \lt\{(4q^2 - a^2 - b^2)^2 + 12a^2b^2 \rt\}^{\frac{3}{2}}
                        \rt]
                     }
                     {\lt\{(4q^2 - a^2 - b^2)^2 - 4a^2b^2 \rt\}^2}.
   \end{array}
\end{equation}
Of course, Eq.(\ref{abqq}) is valid when $\alpha^2 \leq \beta^2 +
\gamma^2 + \delta^2$, where $\{\alpha, \beta, \gamma, \delta \}$
is $\{a, b, q, q \}$ with decreasing order. When $\alpha^2 \geq
\beta^2 + \gamma^2 + \delta^2$, $P_{max}$ will be $\alpha^2 = \max
(a^2, b^2)$.

The dependence of the maximal overlap on state parameters is shown
in Fig.2. The behavior of $P_{max}$ in different limits is
explained in the next section.

\begin{figure}[ht!]
\includegraphics[height=8cm]{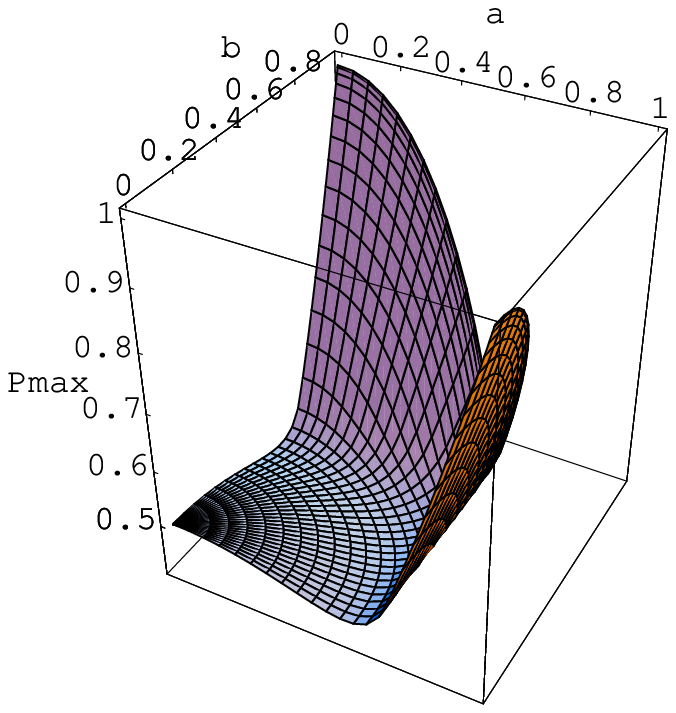}
\caption[fig2]{The maximal overlap $P_{max}$ vs.~the parameters
$a$ and $b$ for the $4$-qubit state. The green and blue areas are
highly entangled regions and the maximal overlap is given by
Eq.(\ref{abqq}). The violet(dark orange) area is a slightly
entangled region and the maximal overlap is $\max(a^2,b^2)$. It is
minimal ($P_{max}=27/64$) at $a=b=1/2$ which is the W-state and
maximal ($P_{max}=1$) either at $a=1, b=0$ or at $a=0,b=1$ which
are product states.}
\end{figure}

\section{Special four-qubit W-type states}

In this section we consider some special $4$-qubit states.

The first one is $a=0$ limit. Since $|W_4\rangle = |0\rangle
\otimes (b|100\rangle + q|010\rangle + q|001\rangle)$ in this
limit, one can compute $P_{max}$ using Eq.(\ref{pmaxw3}). In this
limit Eq.(\ref{abqq}) gives
\begin{equation}
\label{spacial1} P_{max} = \frac{4 q^4}{4q^2 - b^2} \hspace{1,0cm}
(b^2 \leq 2 q^2).
\end{equation}
One can show easily that this is perfectly in agreement with
Eq.(\ref{pmaxw3}).

The second special case is $a=q$ limit. In this limit
Eq.(\ref{abqq}) gives
\begin{equation}
\label{special2} P_{max} = \frac{4 (1-b^2)^3}{(3 - 4 b^2)^2}
\hspace{1,0cm} (b^2 \leq 3 q^2)
\end{equation}
which is also consistent with Eq.(\ref{4-pmax-aaaq}).

The last special case is $2q=a+b$ limit. Although both denominator
and numerator in Eq.(\ref{abqq}) vanish, their ratio has a finite
limit and $P_{max}$ takes correct values in the applicable domain.
The applicable domain is defined by the two restrictions $\alpha^2
\leq \beta^2 + \gamma^2 + \delta^2$ and $2q=a+b$. These
restrictions together with the normalization condition impose
upper and lower bounds for the parameters $a$ and $b$
\begin{equation}\label{lowbound}
\min(a,b)\geq\frac{\sqrt{2}}{6},
\quad\max(a,b)\leq\frac{\sqrt{2}}{2}.
\end{equation}
The maximum overlap probability $P_{max}$ is
\begin{equation}\label{special3}
P_{max}=\frac{27}{256}\frac{(a+b)^4}{ab}.
\end{equation}
The limit $a=b=q=1/2$ again yields $P_{max}$=27/64. Another
interesting limit is the case when $b(a)$ is minimal and $a(b)$ is
maximal. This limit is reached at $a=3b(b=3a)$. Then
Eq.(\ref{special3}) yields $P_{max}=1/2=\alpha^2$. These states
are first type shared states\cite{tama08-1} and allow perfect
teleportation and superdense coding scenario.

\section{discussion}

We have calculated the maximal overlap of one- and two-parametric
W-type states and found their nearest separable states. However,
in some sub-region of the parameter space one can find the nearest
states and corresponding maximal overlaps for generic W-type
states. In fact, the square of any coefficient in
Eq.(\ref{wn-state}) is a maximal overlap in some region of state
parameters. It is easy to check that the product state
$|0_1...0_{k-1}1_k0_{k+1}...0_n\rangle$ is a solution of
stationarity equation with entanglement eigenvalue
$\sqrt{P_{max}}=a_k.$ From previous results one can guess that
this solution gives a true maximum of the overlap if
\begin{equation}\label{sum.ineq}
a_k^2\geq a_1^2+a_2^2+\cdots+a_{k-1}^2+a_{k+1}^2+\cdots+a_n^2 =
1-a_k^2.
\end{equation}
Then the maximal overlap in the slightly entangled region can be
written readily in the form
\begin{equation}\label{sum.pmax}
P_{max}=\max(a_1^2,a_2^2,\cdots,a_n^2)\quad{\rm if}\quad
\max(a_1^2,a_2^2,\cdots,a_n^2)\geq\frac{1}{2}.
\end{equation}
This formula has the following simple interpretation. Equation
(\ref{sum.ineq}) means that the state is already written in the
Schmidt normal form and the maximal overlap takes the value of the
largest coefficient~\cite{gsd-08}.

Now the question at issue is what is happening if
$a_k^2<1/2,\;k=1,2,\cdots,n$. From these inequalities it follows
that
\begin{equation}\label{sum.dis}
\frac{1}{2}\left(a_1+a_2+\cdots+a_n\right)>\max(a_1,a_2,\cdots,a_n).
\end{equation}
From any set of such coefficients one can form
polygons(polyhedrons). This fact is an indirect evidence that
$P_{max}$ has a geometrical meaning. Unfortunately, there is an
obstacle to the goal achievement. The problem is that we have not
the answer for generic states. For example, it is difficult to
conclude from Eq.(\ref{3-pmax-aaq}) that the expression is the
circumradius of a triangle in a particular limit. In general, one
can form many polygons, either convex or crossed, from the set
$a_1, a_2,...,a_n$. Each of them generates its own geometric
quantities that can be treated as the maximal overlap. This
happens because stationarity equations have many solutions in
highly entangled region. And all of these solutions yield the same
expression in particular cases. For example, in
Ref.\cite{tama08-1} it was shown that all convex and crossed
quadrangles are contracted to the same triangle in particular
limits. In conclusion, in order to find a true geometric
interpretation one has to derive $P_{max}$ for generic states.

Another(and probably promising) way to get the desired
interpretation is the following. Since the surface
$(a_1^2-1/2)(a_2^2-1/2)\cdots(a_n^2-1/2)=0$ separates highly and
slightly entangled regions, one may ask what is happening on this
surface. That is, we are considering polygons whose sides satisfy
the equality $a_k^2 = a_1^2 + a_2^2 + \cdots + a_{k-1}^2 +
a_{k+1}^2+\cdots+a_n^2$ for any $k$. For $n=3$ we perfectly know
that corresponding polygons are right triangles and the center of
a circumcircle lies on the largest side of a right triangle. Then,
we can conclude that if the center of the circumcircle is inside
the triangle, then the maximal overlap is the circumradius and
otherwise is the largest coefficient. However, for $n\geq4$ we do
not know what are the polygons for which the square of the largest
side is the sum of squares of the remaining coefficients. If one
understands the geometric meaning of this relation, then one finds
a clue. And this clue may enable us to find $P_{max}$ for generic
W-type states. These type of analytic expressions can have
practical application in QICC and may shed new light on
multipartite entanglement.

All above-mentioned problems owe their origin to the fact that the
injective tensor norm is related to the Cayley's
Hyperdeterminant~\cite{sud-geom}. It is well-known that this
hyperdeterminant has a geometrical interpretation for $n=3$ and no
such interpretation is known for $n\geq4$ so far. We hope to keep
on studying this issue in the future.

\medskip

\begin{acknowledgments}
This work was supported by the Kyungnam University Foundation,
2008.
\end{acknowledgments}

\bigskip


\end{document}